\documentclass[12pt]{article}
\usepackage{epsfig}
\newcommand{\beq}{\begin{eqnarray}}
\newcommand{\eeq}{\end{eqnarray}}
\newcommand{\half}{\frac12}
\newcommand{\bi}{\bibitem}
\newcommand{\wg}{\wedge}
\def\gsim{\,\lower2truept\hbox{${> \atop\hbox{\raise3truept\hbox{$\sim$}}}$\,}}

\begin{document}
\baselineskip=15.5pt

\begin{titlepage}

\begin{flushright}
hep-th/0212228\\
PUPT-2071\\
\end{flushright}
\vfil

\begin{center}
{\huge Semiclassical Quantization of Giant Gravitons}\\
\vspace{3mm}
\end{center}

\vfil
\begin{center}
{\large Peter Ouyang }\\
\vspace{1mm}
Joseph Henry Laboratories, Princeton University,\\
Princeton, New Jersey 08544, USA\\
\vspace{3mm}
\end{center}

\vfil

\begin{center}
{\large Abstract}
\end{center}
We study excited spherical branes (``giant gravitons'') in $AdS \times
S$ spacetimes with background flux.  For large excitation, these
branes may be treated semiclassically.  We compute their spectra using
Bohr-Sommerfeld quantization and use the AdS/CFT correspondence to
relate them to anomalous dimensions in the dual field theory at strong
coupling, expressed as a series expansion in powers of $1/N$.  These
effects resemble those due to $k$-body forces between quarks in
Hartree-Fock models of baryons at large $N$.  For branes expanded in
AdS, we argue that the anomalous dimensions are due to loop
corrections to the effective action.

\noindent

\vfil
\begin{flushleft}
December 2002
\end{flushleft}
\vfil
\end{titlepage}
\newpage

\section{Introduction}

The AdS/CFT correspondence \cite{jthroat,US,EW} relates weakly coupled
string theories in backgrounds with flux and curvature to strongly
coupled field theories. This duality is well-understood in the the
string background is nearly flat, but our understanding is more
rudimentary in highly curved backgrounds; in the former case the
massive string modes decouple and we can work with a low-energy
effective theory, whereas in the latter case string effects are often
non-negligible.

One approach to studying nonzero curvature relies on semiclassical
methods \cite{GKP}.  Highly excited strings can probe length scales on
the order of the radius of curvature of the ambient spacetime, which
normally makes calculations difficult.  However, in the limit of high
excitation semiclassical methods are reliable and, at least for some
states of high symmetry, it turns out that we can extract some
physical insight without working with the full quantum string theory.
The authors of \cite{GKP} worked primarily with ``folded'' closed
strings, but also remarked on the existence of pulsating string
solutions, which \cite{Minahan} subsequently studied using
Bohr-Sommerfeld techniques.

In this paper, we present some investigations on the effects of finite
curvature for excited spinning spherical brane probes.  Unlike the
case of F-strings studied in \cite{GKP, Minahan}, these curvature
effects introduce $1/N$ corrections rather than $g_sN$ corrections to
operator dimensions in the field theory.  Brane probes also differ
crucially from the previously studied strings in that they couple
non-trivially to the background RR flux filling the spacetime, as
observed by \cite{mms,d1d5,giants,grisaru, hashimoto}.  Interaction
with the background flux causes the branes to expand to finite size;
these expanded branes are supersymmetric, and are known in the
literature as ``giant gravitons.'' Some small fluctuations about these
BPS states have been studied in earlier work
\cite{das,mikhailov,kim,takayanagi}; we will compute the dimensions of
the operators corresponding to these states when the fluctuations are
not small.  In Section 2, we present the case of spherical D3-branes
in $AdS_5 \times S^5$.  We briefly remark on the case of $AdS_3 \times
S^3$ in Section 3, for which our techniques do not apply.  In Section
4, we give the analogous results for the M-theory solutions
$AdS_4\times S^7$ and $AdS_7\times S^4$, whose field theory duals are
poorly understood.  Finally, we comment on the range of validity and
physical interpretation of our results in Section 5.

\section{D3-Branes in $AdS_5\times S^5$}

Let us begin by studying oscillating branes in $AdS_5\times S^5$. Our
convention for the full ten-dimensional metric is
\beq
ds^2= L^2\left(-\cosh^2\rho dt^2 + d\rho^2 +\sinh^2\rho d\Omega_3^{~2}+\cos^2\theta d\psi^2 + d\theta^2 +\sin^2 \theta d\Omega_{3'}^{~2}\right)
\label{metric}
\eeq
where we have chosen global coordinates for the $AdS_5$.  The RR
four-form potential is
\beq
C_4 = \frac{L^4}{g_s} \left( \sinh^4\rho dt \wg \Omega_3-\sin^4\theta d\psi \wg \Omega_{3'} \right)
\eeq
and all the other supergravity fields vanish.  The AdS radius is given by $L^4=4\pi g_s N \alpha'^2$.

\subsection{Oscillating D3-Branes in $AdS_5$}

A simple way to embed an oscillating D3-brane in this geometry is to
wrap the brane on the 3-sphere, and to let the radial position be a
function of coordinate time on the brane, $\rho=\rho(\tau)$.  If we
also introduce angular momentum on the $S^5$ by making $\psi$ a
function of $\tau$, then the action, consisting of the
Dirac-Born-Infeld and Wess-Zumino terms, is
\beq
S=-T_3 L^4 \int \Omega_3 \int d\tau \left( \sinh^3 \rho \sqrt{\dot{t}^2 \cosh^2\rho -\dot{\rho}^2-\dot{\psi}^2} - \sigma \dot{t} \sinh^4 \rho \right).
\label{d3action}
\eeq
In this form, AdS coordinate time is $t=t(\tau)$.  We also introduce
$\sigma = \pm 1$, with the upper sign corresponding to a D3-brane and
the lower to an anti-D3.  Notice that the coefficient of the action is
simply
\beq
T_3 L^4 \int \Omega_3 = N.
\eeq
For the semiclassical calculation we perform, the relevant quantity is
really the action in $\hbar$ units: $S/\hbar \sim N/\hbar$.  Thus,
from the standpoint of the D-brane action (we ignore fluctuations of
supergravity fields in the bulk of AdS), the semiclassical expansion
is a series in powers of $\hbar/N$.  We will set $\hbar =1$ in the
rest of this paper and in this section take the equivalent $N\gg 1$ limit.

Now let us find the classical equations of motion from this action.
The Lagrangian does not depend on $t$, so the quantity
\beq
\frac{\partial L}{\partial \dot{t}} = 
-N\left(\frac{\dot{t}\sinh^3\rho \cosh^2 \rho}{\sqrt{\dot{t}^2 \cosh^2\rho -\dot{\rho}^2-\dot{\psi}^2}} -\sigma \sinh^4 \rho\right) \equiv -E
\eeq
is conserved.  In terms of the integration constant $E$, we have
\beq
(-\dot{\rho}^2 -\dot{\psi}^2+\dot{t}^2 \cosh^2 \rho)(E/N+\sigma \sinh^4\rho)^2 = \dot{t}^2 \cosh^4 \rho \sinh^6\rho.
\eeq
A convenient gauge choice is
\beq
\dot{t}^2 \cosh^4\rho =  (E/N +\sigma \sinh^4\rho)^2.
\label{gauge}
\eeq
{}From the $\psi$ equation of motion, and using the gauge choice
(\ref{gauge}), we identify the conserved angular momentum
\beq
\frac{\partial{L}}{\partial{\dot{\psi}}} = N\frac{\dot{\psi}\sinh^3\rho}{\sqrt{\dot{t}^2 \cosh^2\rho -\dot{\rho}^2-\dot{\psi}^2}}=N\dot{\psi} \equiv J
\eeq
which allows us to write the once-integrated equation of motion for
$\rho$ as
\beq
\dot{\rho}^2 = \frac{(E/N+\sigma  \sinh^4\rho)^2}{\cosh^2\rho} -\sinh^6\rho-\left(\frac{J}{N}\right)^2.
\label{rhodot}
\eeq
Note that this equation of motion corresponds to one-dimensional motion of a particle in a potential 
\beq
V(\rho)=\frac1{2\cosh^2\rho}\left( 2\sigma \frac{E}{N}\sinh^4\rho-\sinh^6\rho-\left(\frac{E}{N}\right)^2\sinh^2\rho  \right)
\label{potential}
\eeq
with energy $\half \left(\frac{E-J}{N}\right)^2$. It is instructive to
consider the unintegrated equation of motion for $\rho$, which we can
obtain by differentiating equation (\ref{rhodot}) with respect to
$\tau$:
\beq
\ddot{\rho} = -\frac{\sinh\rho}{\cosh^3\rho}(E/N+\sigma \sinh^4\rho)^2+2\sigma\frac{\sinh^3\rho}{\cosh\rho}(E/N+\sigma \sinh^4\rho)-3\sinh^5\rho \cosh\rho.
\label{rhoddot}
\eeq
The first and last terms in equation (\ref{rhoddot}) correspond to
forces directed radially inward, while the middle term gives an
outward force for $\sigma=1$ (as it should, for a D3-brane) and an
inward force for $\sigma = -1$ (for an anti-brane.)

When $E=J$, there are special stable states.  The equation of motion becomes
\beq
\dot{\rho}^2 = -\frac{\sinh^2\rho}{\cosh^2\rho} \left(\frac{E}{N} -\sigma \sinh^2\rho\right)^2
\eeq
so that both $\dot{\rho}$ and $\ddot{\rho}$ vanish at $\rho=0$ and,
for $\sigma=+1$, at $\rho=\sinh^{-1} \sqrt{\frac{E}{N}}$.  These
static spherical branes are the BPS ``dual giant gravitons''
discovered in \cite{grisaru,hashimoto}.  Whether or not the brane at
$\rho=0$ is a physically sensible state is not obvious -- the brane
becomes highly curved and we should not trust the DBI action.  We will
only consider the expanded branes, and will always take
\beq
E\gsim N
\eeq
so that the size of the D3-brane remains finite as we take $N \gg 1$.  For the remainder of this subsection, $\sigma = +1$.

The frequency of small oscillations about the expanded BPS brane state
is relatively simple to calculate.  This computation was previously
carried out in \cite{das}; we will reproduce their result with a
slightly different method.  Making the convenient change of variables
$y=\sinh\rho$, equation (\ref{rhodot}) becomes
\beq
\dot{y}^2 = \left(\frac{E}{N}\right)^2-\left(\frac{J}{N}\right)^2 (y^2+1) +2 \frac{E}{N} y^4 -y^6
\label{ydot}
\eeq
with static brane solutions at $y=0$ and $y^2=E/N$.  To find the
frequency of oscillations about the latter solution, we set $J=E$ and
treat the left hand side of (\ref{ydot}) as an
inverted potential $-2U(y)$.  The frequency of small oscillations is
then
\beq
\Omega^2 = U''(y=\sqrt{\frac{E}{N}})= 4\left(\frac{E}{N}\right)^2
\eeq
This agrees with the result of \cite{das} for spherical branes,
provided that we convert our result to static gauge ($t=\tau$).

Now let us analyze these excited giant gravitons using semiclassical
methods.  The basic approach is to use Bohr-Sommerfeld quantization,
as was done for oscillating strings by \cite{Minahan}.  The momentum
conjugate to $\rho$ is
\beq
P_{\rho} &=& N\frac{\sinh^3\rho \dot{\rho}}{\sqrt{\dot{t}^2 \cosh^2\rho -\dot{\rho}^2-\dot{\psi}^2}}=N\dot{\rho} \\
&=& N\dot{\rho}
\eeq
One period consists of the expansion of the brane from a minimum
radius $\rho_1$ to its maximum radius $\rho_2$ and the subsequent
contraction.  The Bohr-Sommerfeld condition, which will be a good
approximation for $n\gg 1$, is
\beq
2\pi n &=& \oint P_{\rho} d\rho =2\int_{\rho_1}^{\rho_2} N \dot{\rho} d\rho \\
& &\hspace{-1.5cm}= 2N\int_{\rho_1}^{\rho_2}\frac{d\rho}{\cosh\rho} \sqrt{\left(\frac{E}{N}\right)^2 -\left(\frac{J}{N}\right)^2 +2 \frac{E}{N}\sinh^4\rho-\sinh^6\rho-\left(\frac{J}{N}\right)^2\sinh^2\rho}.
\label{wkbint}
\eeq
This integral can be expressed in terms of highly unenlightening
elliptic integrals.  We can learn much more in the case where
the giant gravitons are not too strongly excited:
\beq
E-J \ll J, N.
\eeq
Now let us define $x=\sinh^2\rho$.  Inside the square root, we find a
cubic polynomial.  For the trajectories of interest, there are three
real roots, which we call $a,b_1,b_2$.  The integral above becomes
\beq
N\int_{b_1}^{b_2} \frac{dx}{x^{1/2} (1+x)} \sqrt{(x-a)(x-b_1)(b_2-x)}
\eeq
and for small $\Delta\equiv (E-J)/N$, we can expand the roots as
\beq
a &\simeq& 2\frac{\Delta}{\omega} + \frac{8 +\omega}{\omega^3}\Delta^2 \\
b_1 &\simeq& \omega -\sqrt{2\left(1+\omega\right)}\Delta^{1/2} + \frac{\omega-1}{\omega} \Delta-\frac{5+3\omega+\omega^2}{2^{3/2} \left(1+\omega\right)^{1/2}\omega^2}\Delta^{3/2} -\frac{8+\omega}{2\omega^3} \Delta^2 \\
b_2 &\simeq& \omega +\sqrt{2\left(1+\omega\right)}\Delta^{1/2} + \frac{\omega-1}{\omega} \Delta+\frac{5+3\omega+\omega^2}{2^{3/2} \left(1+\omega\right)^{1/2}\omega^2}\Delta^{3/2}-\frac{8+\omega}{2\omega^3} \Delta^2 
\eeq
where we have defined $\omega=J/N$.  To evaluate the integral
perturbatively in $\Delta$, we make a change of variables:
\beq
x \equiv \left( \frac{b_2-b_1}{2} \right) \xi + \left( \frac{b_2+b_1}{2} \right).
\eeq
Then the action integral becomes
\beq
2\pi n = N \left( \frac{b_2-b_1}{2} \right)^2 \int_{-1}^{1} d\xi \sqrt{1-\xi^2} \left[\frac{1}{1+\omega}+\Delta \left(\frac{2\xi^2}{(1+\omega)^2}-\frac{1+\omega^{-2}}{(1+\omega)^2}\right) \right]
\eeq
Evaluating the integrals, the quantization condition is, at quadratic
order in $\Delta$,
\beq
2n = N\left(\Delta + \frac{3}{2\omega^2}\Delta^2 \right)
\eeq
or, solving for $E-J$,
\beq
E-J = 2n -\frac{6n^2N}{J^2}.
\label{specads5}
\eeq
In the AdS/CFT correspondence, we identify the energy $E$ in global
coordinates as the operator dimension in the field theory and $J$ as
the R-charge, so we have computed the spectrum of dimensions of
operators corresponding to these excited branes.  We will comment
further on this result in section 5.

Our analysis breaks down if the D3-brane becomes too highly excited.
Consider the one-dimensional effective potential in terms of the
coordinate $\rho$, as determined from Eq. (\ref{rhodot}), which we
show in Figure 1.
\begin{figure}[htb]
\begin{center}
\epsfig{figure=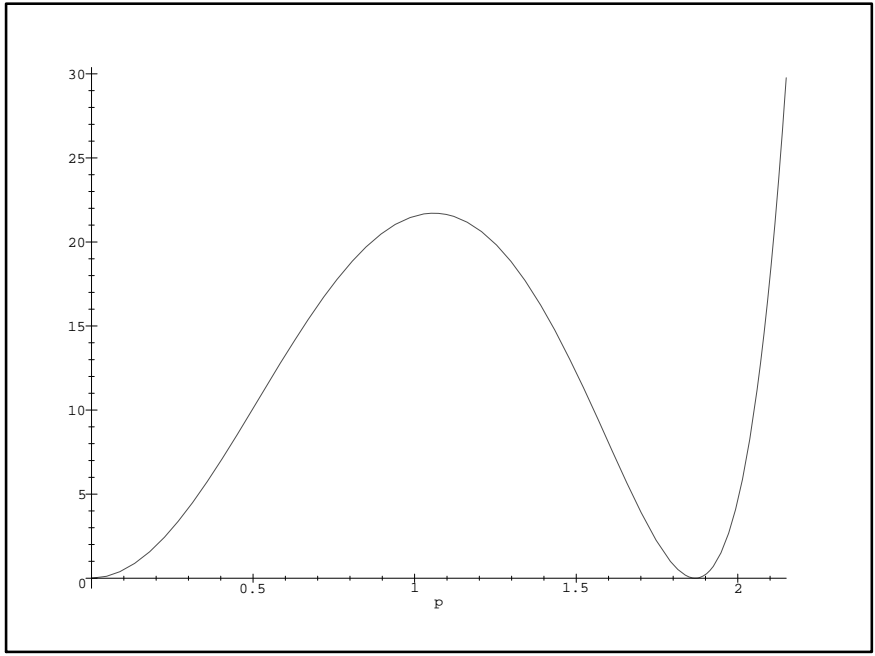,angle=270,width=5in}
\caption{$V(\rho)$, for $E/N=10$}
\end{center}
\label{vrho01}
\end{figure} 
For small $E-J$, the oscillations are small and we can consider the
solutions which are always at large radius.  However, if $E-J$ is too
large, then the amplitude of oscillations can be large enough to
excite the D3-brane over the potential barrier in Figure 1, and the
brane will collapse to small radius.  As mentioned earlier, at small
$\rho$ the brane is highly curved and we can no longer rely on the DBI
action.

For large energy, there is a substantial regime of validity of our
calculation.  In this limit, it is simple to determine the minimum
R-charge $J$ compatible with stable oscillations -- we simply take the
zeros of $V'$, computed from (\ref{potential}) and ask whether
$\dot{\rho}^2$ is positive, taking $J\ll E$.  The result is that if
\beq
\frac{J}{N} > \sqrt{2} \left(\frac{E}{N}\right)^{3/4}
\eeq
then the oscillations are constrained to large radius. 

\subsection{Oscillating D3-Branes on $S^5$}

The original giant gravitons of \cite{giants} were D3-branes wrapped
on an $S^3$ in the $S^5$.  They also move along a circle in the $S^5$
and are thus coupled to the background RR five-form flux.  The $S^3$
on which these 3-branes are wrapped is contractible, but the RR-flux
exerts a repulsive force on the brane, stabilizing it at finite size
on the $S^5$.  We will now study excitations of these branes.

Using the spacetime geometry of (\ref{metric}), we take the embedding
$t=t(\tau)$, $\psi=\psi(\tau)$, and $\theta=$ constant.  The D3-brane
action becomes
\beq
S=-N\int d\tau \left( \sin^3 \theta \sqrt{\dot{t}^2-\cos^2\theta \dot{\psi}^2-\dot{\theta}^2} +\sigma \sin^4 \theta \dot{\psi} \right)
\eeq
We choose the gauge 
\beq
\dot{t}^2-\cos^2\theta \dot{\psi}^2 -\dot{\theta}^2 = \sin^{6} \theta
\eeq
and define the conserved quantities
\beq
E&=&-\frac{\partial L}{\partial \dot{t}}=N\sqrt{\dot{\theta}^2+\cos^2\theta\dot{\psi}^2+\sin^{2p}\theta}=N\dot{t}, \\
J &=&N\omega=\frac{\partial L}{\partial \dot{\psi}}= N\left(\dot{\psi}\cos^2\theta - \sigma\sin^4\theta \right)
\eeq
in terms of which we may write the momentum conjugate to $\theta$,
\beq
P_{\theta} = N\dot{\theta}=N\sqrt{\left(\frac{E}{N}\right)^2-\frac{(\omega+\sigma \sin^4\theta)^2}{\cos^2\theta}-\sin^6 \theta}.
\eeq
The static ``giant graviton'' solutions exist for $\sigma = -1$, $\omega = E/N$, and
$\omega \le 1$.

The Bohr-Sommerfeld quantization condition may be analyzed in much the
same way as in the previous subsection.  Though the expressions differ
slightly in intermediate steps, the final result is the same:
\beq
E-J \simeq 2n -\frac{6n^2N}{J^2}.
\label{spherespec}
\eeq

\section{Remark on Long Strings in $AdS_3$}

The oscillating string solutions have already been thoroughly studied
in the case of $AdS_3 \times S^3$ \cite{mms,d1d5,mo}, so we will only
discuss them briefly here.  The action for a string coupled to the
background flux is
\beq
S=-\eta \int d\tau \left( \sinh \rho \sqrt{\dot{t}^2 \cosh^2\rho -\dot{\rho}^2-\dot{\psi}^2} - \sigma \dot{t} \sinh^2 \rho \right)
\eeq
where $\eta = 2\pi TL^2$.  We again identify the conserved energy as
$E=-\frac{\partial L}{\partial \dot{t}}$ and the angular velocity
$\omega=\dot{\psi}$, and choose the gauge
\beq
\dot{t}^2 \cosh^4\rho =  (E/\eta +\sigma \sinh^{2}\rho)^2.
\eeq
Then we find that $P_{\rho} = \eta \dot{\rho}$ and that
\beq
\dot{\rho}^2&=&\frac{(E/\eta + \sigma \sinh^{2}\rho)^2}{\cosh^2 \rho}-\sinh^{2}\rho -\omega^2\\
&=& \frac{1}{\cosh^2\rho} \left((E/\eta)^2-\omega^2 -(\omega^2-2\sigma E/\eta +1)\sinh^2\rho\right).
\label{rhodotads3}
\eeq
Notice that if $\omega=0$ and
\beq
\frac{2\sigma E}{\eta} \ge 1
\eeq
then there is no value of $\rho$ for which $\dot{\rho}$ vanishes.
These solutions, for which there is no turning point of the motion,
are the so-called ``long strings'' in $AdS_3$.  For nonzero $\omega$,
there are no giant graviton solutions; from Eq.(\ref{rhodotads3}) we
can see that there is at most one point at which $\dot{\rho}=0$.  Thus
for large energy the radial motion cannot be periodic, and the
Bohr-Sommerfeld analysis of Section 2 is inapplicable.

\section{Oscillating Branes in $AdS_4 \times S^7$ and $AdS_7\times S^4$}

Calculations similar to those in Section 2 can be done for AdS spaces
of dimensions other than 5.  In this section we examine the cases of
special interest, $AdS_4 \times S^7$ and $AdS_7\times S^4$, which are
solutions of M-theory.  The calculations are very similar to those for
$AdS_5 \times S^5$, so we will list our results first; the details
follow for interested readers.  For M2-branes oscillating in $AdS_4$,
\beq
E-\frac{J}{2} = n -12 \frac{Nn^2}{J^3}, \qquad J\sim N^{1/2}.
\label{specads4}
\eeq
For M5-branes in $S^7$,
\beq
E-\frac{J}{2}=2n-\frac{15n^2N^{1/2}}{2J^{3/2}}, \qquad J\sim N.
\label{specs7}
\eeq
For M5-branes oscillating in $AdS_7$,
\beq
E-2J=4n-\frac{15Nn^2}{2J^{3/2}}, \qquad J\sim N^2.
\label{specads7}
\eeq
For M2-branes in $S^4$,
\beq
E-2J=2n-\frac{6N^2n^2}{J^3},\qquad J\sim N .
\label{specs4}
\eeq

At $n=0$, these calculations reproduce the BPS bounds for the
conformal field theories on the boundaries of $AdS_4$ and $AdS_7$,
while at nonzero $n$ the leading terms are consistent with a
semiclassical calculation in field theory.  We will comment further on
these results in Section 5.

\subsection{Oscillating Membranes in $AdS_4$}

The eleven-dimensional metric is
\beq
ds^2= L^2\left(-\cosh^2\rho dt^2 + d\rho^2 +\sinh^2\rho d\Omega_2^{~2}\right)+4L^2\left(\cos^2\theta
d\psi^2 + d\theta^2 +\sin^2 \theta d\Omega_{5}^{~2}\right)
\label{metric47}
\eeq
where we have chosen global coordinates for the $AdS_4$.  The
three-form potential of M-theory is
\beq
A_3 = L^3\sinh^3\rho dt \wg \Omega_2
\eeq

The action for an M2-brane wrapped on the $S^2$ and spinning in the
$S^7$ is
\beq
S=-T_{M2} L^{3} Vol({\bf S^2}) \int d\tau \left( \sinh^2 \rho \sqrt{\dot{t}^2 \cosh^2\rho -\dot{\rho}^2-4\dot{\psi}^2} - \sigma \dot{t} \sinh^{3} \rho \right).
\label{m2action}
\eeq
Recall that the tension of an M2-brane is $T_{M2}= (2\pi)^{-2}
l_p^{~-3}$ and that the $AdS_4$ radius for the $AdS_4 \times S^7$
solution of M-theory is given by $L_{AdS_4}= \left(\frac{\pi^2
N}{2}\right)^{1/6} l_p$.  The coefficient of the action is therefore
$\sqrt{\frac{N}{2}}$.  The conserved energy is
\beq
\frac{\partial L}{\partial \dot{t}} =
-\sqrt{\frac{N}{2}}\left(\frac{\dot{t}\sinh^2\rho\cosh^2\rho}{\sqrt{\dot{t}^2 \cosh^2\rho -\dot{\rho}^2-4\dot{\psi}^2}} -\sigma \sinh^{3} \rho\right) \equiv -E
\eeq
and the conserved angular momentum is
\beq
\frac{\partial L}{\partial \dot{\psi}} =
\sqrt{\frac{N}{2}}\left(\frac{4\dot{\psi}\sinh^2\rho}{\sqrt{\dot{t}^2 \cosh^2\rho -\dot{\rho}^2-4\dot{\psi}^2}} \right) \equiv J.
\eeq
We define $c=E\sqrt{\frac{2}{N}}$ and $4\omega=J\sqrt{\frac{2}{N}}$.
With the gauge choice
\beq
\dot{t}^2 \cosh^4\rho =  \left(c +\sigma \sinh^3\rho\right)^2
\eeq
we obtain $\dot{\psi}=\omega$, and the equation of motion for the radial position $\rho$ is
\beq
\dot{\rho}^2 = \frac{(c +\sigma \sinh^3\rho)^2}{\cosh^2\rho}-\sinh^4\rho-4\omega^2
\eeq
When $c=2\omega$ and $\sigma = +1$ there are special static solutions.
The momentum conjugate to $\rho$ is $P_{\rho}
=\sqrt{\frac{N}{2}}\dot{\rho}$.  By a similar analysis as in the
previous section, we find that to quadratic order in
$\Delta=c-2\omega$, is that
\beq
n= \sqrt{\frac{N}{2}}\left(\Delta + \frac{3}{8\omega^3} \Delta^2 \right).
\eeq
As a condition on allowed energies, we find
\beq
E-\frac{J}{2} = n -12 \frac{Nn^2}{J^3}
\eeq

\subsection{Oscillating M5-branes in $AdS_4\times S^7$}

In the $AdS_4\times S^7$ solution, we can also embed an M5-brane by
wrapping it on a 5-sphere in the $S^7$.  The M5 also spins in the
$\psi$-direction, as usual, and expands to an angle $\theta$.  The
relevant six-form potential of M-theory is (using the requirement
$F_7=\star F_4$)
\beq
A_6 = -(2L)^6 \sin^6\theta d\psi \wg \Omega_5
\eeq
As in the previous subsection, the $AdS_4$ radius $L$ is given by
$L_{AdS_4}= \left(\frac{\pi^2 N}{2}\right)^{1/6} l_p$.  The $M5$
action is
\beq
S=-32T_{M5} L^{6} Vol({\bf S^5}) \int d\tau \left( \sin^5 \theta \sqrt{\dot{t}^2 \cosh^2\rho -4\dot{\theta}^2-4\cos^2\theta \dot{\psi}^2} - 2\sigma \dot{\psi} \sin^{6} \theta \right).
\label{m5s7action}
\eeq
Because the tension of an M5-brane is $T_{M5}= (2\pi)^{-5} l_p^{~-6}$,
we find that the coefficient in front of the action integral is
$\frac{N}{2}$.  The conserved quantities $E$ and $J$ are given by
\beq
\frac{\partial L}{\partial \dot{t}} =
-\frac{N}{2}\left(\frac{\dot{t}\sin^5\theta}{\sqrt{\dot{t}^2 -4\dot{\theta}^2-4\cos^2\theta \dot{\psi}^2}}\right) \equiv -E
\eeq
and
\beq
\frac{\partial L}{\partial \dot{\psi}} =
2N\frac{\dot{\psi}\cos^2\theta\sin^5\theta}{\sqrt{\dot{t}^2 -4\dot{\theta}^2-4\cos^2\theta\dot{\psi}^2}} -N\sigma\sin^6\theta \equiv J.
\eeq
The most convenient gauge choice is
\beq
\sqrt{\dot{t}^2 -4\dot{\theta}^2-4\cos^2\theta\dot{\psi}^2} =\sin^5\theta
\eeq
so that $E=\frac{N}{2} \dot{t}$ and $J= 2N\dot{\psi} \cos^2\theta$.
Moreover, the momentum conjugate to $\theta$ is
\beq
P_{\theta} &=& 2N\dot{\theta}  \\
&=& \frac{N}{\cos\theta} \sqrt{\left(\frac{2E}{N}\right)^2-\left(\frac{J}{N}\right)^2-\left(\frac{2E}{N}\right)^2 \sin^2 \theta-2\frac{J}{N}\sigma \sin^6\theta-\sin^{10}\theta}
\eeq
We see that the giant graviton solutions will have $\sigma = -1$.  Computing the Bohr-Sommerfeld integral as before,
\beq
2\pi n = \oint P_{\theta} d\theta
\eeq
where now we work perturbatively in $(2E-J)/N$, we obtain the spectrum 
\beq
E-\frac{J}{2}=2n-\frac{15n^2N^{1/2}}{2J^{3/2}}.
\eeq

\subsection{Oscillating M5-branes in $AdS_7 \times S^4$}

The eleven-dimensional metric is
\beq
ds^2= L^2\left(-\cosh^2\rho dt^2 + d\rho^2 +\sinh^2\rho d\Omega_2^{~2}\right)+\frac{L^2}{4}\left(\cos^2\theta
d\psi^2 + d\theta^2 +\sin^2 \theta d\Omega_{5}^{~2}\right)
\label{metric74}
\eeq
where we have chosen global coordinates for the $AdS_7$.  The
six-form potential of M-theory is
\beq
A_6 = -L^6 \sinh^6\rho dt \wg \Omega_5
\eeq

The action for an M5-brane wrapped on the $S^2$ and spinning in the
$S^7$ is
\beq
S=-T_{M5} L^{6} Vol({\bf S^5}) \int d\tau \left( \sinh^5 \rho \sqrt{\dot{t}^2 \cosh^2\rho -\dot{\rho}^2-\frac14\dot{\psi}^2} + \sigma \dot{t} \sinh^{6} \rho \right).
\label{m5action}
\eeq
Recall that the tension of an M5-brane is $T_{M5}= (2\pi)^{-5}
l_p^{~-6}$ and that the $AdS_7$ radius for the $AdS_7 \times S^4$
solution of M-theory is given by $L_{AdS_7}= 2\left(\pi N\right)^{1/3}
l_p$.  The coefficient of the action is therefore $2N^2$.  The
conserved energy is
\beq
\frac{\partial L}{\partial \dot{t}} =
-2N^2\left(\frac{\dot{t}\sinh^5\rho\cosh^2\rho}{\sqrt{\dot{t}^2 \cosh^2\rho -\dot{\rho}^2-\frac14 \dot{\psi}^2}} +\sigma \sinh^{6} \rho\right) \equiv -E
\eeq
and the conserved angular momentum is
\beq
\frac{\partial L}{\partial \dot{\psi}} =
\frac{N^2}{2}\left(\frac{\dot{\psi}\sinh^5\rho}{\sqrt{\dot{t}^2 \cosh^2\rho -\dot{\rho}^2-\frac14\dot{\psi}^2}} \right) \equiv J.
\eeq
We define $c=E/2N^2$ and $\omega=2J/N^2$.  With the gauge choice
\beq
\dot{t}^2 \cosh^4\rho =  \left(c -\sigma \sinh^6\rho\right)^2
\eeq
we obtain $\dot{\psi}=\omega$, and the equation of motion for the
radial position $\rho$ is
\beq
\dot{\rho}^2 = \frac{(c -\sigma \sinh^6\rho)^2}{\cosh^2\rho}-\sinh^{10}\rho-\frac{\omega^2}{4}
\eeq
Here, we take $c \simeq \omega/2$ and $\sigma = -1$ to study
oscillations about the special static solutions.  The momentum
conjugate to $\rho$ is $P_{\rho} = 2N^2\dot{\rho}$.  Computing the
integrals as before, we obtain
\beq
E-2J = 4n -\frac{15Nn^2}{2J^{3/2}}.
\eeq

\subsection{Oscillating M2 branes in $AdS_7\times S^4$}

In the eleven-dimensional geometry of the previous subsection, we may
also study giant gravitons which are M2-branes wrapped on a two-sphere
embedded in the $S^4$.  The 3-form potential is
\beq
A_3 = \left(\frac{L}{2}\right)^3\sin^3\theta d\psi \wg \Omega_2
\eeq
The action for an M2-brane wrapped on the $S^2$ and spinning in the
$S^4$ (along an angular coordinate $\psi$) is
\beq
S=-\frac18 T_{M2} L^{3} Vol({\bf S^2}) \int d\tau \left( 2\sin^2 \theta \sqrt{\dot{t}^2 -\frac{\dot{\theta}^2}{4}-\frac{\dot{\psi}^2\cos^2\theta}{4}} - \sigma \dot{\psi} \sin^{3} \theta \right).
\label{m2actions4}
\eeq
The tension of an M2-brane is $T_{M2}= (2\pi)^{-2}
l_p^{~-3}$ and the $AdS_7$ radius is given by $L= 2\left(\pi N\right)^{1/3}
l_p$, so the coefficient of the action is just $N$.  In the gauge
\beq
\sqrt{\dot{t}^2 -\frac{\dot{\theta}^2}{4}-\frac{\dot{\psi}^2\cos^2\theta}{4}} =\sin^2\theta
\eeq
we have
\beq
\frac{\partial L}{\partial \dot{\psi}} &=& J =\frac{N}{2}\dot{\psi} \cos^2\theta-\sigma N \sin^3\theta \\
-\frac{\partial L}{\partial \dot{t}} &=& E = 2N\dot{t} \\
\frac{\partial L}{\partial \dot{\theta}} &=& P_{\theta} = \frac{N}{2}\dot{\theta}.
\eeq
Now, we take $\sigma = -1$ and work perturbatively in $\frac{E}{2N} -\frac{J}{N}$.  Evaluating the Bohr-Sommerfeld integral in this case yields
\beq
E-2J=2n-\frac{6N^2n^2}{J^3}.
\eeq

\section{Comments and Interpretation}

Our results are easiest to interpret for the oscillating branes on
$S^5$.  For these states, there is a conjecture that the dual field
theory operators are subdeterminants \cite{bbns} of the form
\beq
Z_{i_1}^{~j_1} \cdots Z_{i_J}^{~j_J} \epsilon^{i_1 \cdots i_J i_{J+1} \cdots i_N} \epsilon_{j_1\cdots j_J i_{J+1} \cdots i_N}
\label{subdet}
\eeq
where the summed indices are $SU(N)$ gauge indices.  Subsequent work
supporting this conjecture includes \cite{corley, aharony}. Following
\cite{bhk}, non-BPS excitations of this giant graviton should
correspond to insertions of fields into the operator (\ref{subdet}).
These insertions ought to involve scalars $\phi$ and possibly
derivatives $\nabla_{\mu}$ (fermions and gauge fields are associated
with scalar and two-form potentials in the string background.)  For
fixed R-charge, any $\phi$ must actually appear in the combination
$\phi \bar{\phi}$, and Poincar\'{e} symmetry of the background
requires the derivative to be contracted into something -- perhaps
$\nabla^2$. Then from the usual dimension counting in a
four-dimensional field theory, the quantity $E-J$ must change in units
of two.

The interesting physics arises from the corrections to the leading
behavior, which we can interpret as arising from many-body forces on
the field theory side.  Consider replacing some of the $Z_i^{~j}$
fields in the operator (\ref{subdet}) by $(Z\bar{\phi}\phi)_i^{~j}$
(we might also imagine taking two $Z$ fields and replacing one by
$Z\phi$ and the other by $Z\bar{\phi}$.)  Excitations of this form
have been conjectured to correspond to open string states
\cite{bhk,bhln,takayanagi}, which can change the size and shape of the
D3-brane.  This insertion of ``impurities'' into the subdeterminant
operator is reminiscent of a Hartree-Fock analysis of baryons
\cite{wittenbar}, in which we excite individual quarks.  The
correction term, which we found to be of order $n^2/N$, then
corresponds to two-body interactions between impurities (the two-body
force is of order $1/N$, and there are roughly $n^2$ ways to pick two
impurities.)  The minus sign indicates that the force between excited
quarks is attractive.  Note that if we had computed the spectrum to
higher orders in $\Delta$ in Section 2, we would have obtained terms
of order $n^k/N^{k-1}$.  Assuming that the semiclassical expansion
remains valid for these terms, we can thus continue the expansions
peformed in this paper to obtain leading contributions to the $k$-body
forces for any $k$.

The results for M-branes which have expanded on the sphere are more
difficult to interpret because we lack a Lagrangian description of the
dual field theories.  However, the corrections in the semiclassical
expansion again have the form $n^k/N^{k-1}$, suggesting that the
picture of having $k$-body forces between quarks extends to the
M-brane case.

Similar results have appeared previously \cite{bhln,takayanagi} in the
context of the plane-wave approximation of $AdS_5 \times S^5$
\cite{BMN}.  We have performed our calculations in a different regime
-- we take $J \sim N$, as opposed to the standard plane-wave limit,
for which $J^2/N$ is held fixed.  Note that the corrections to the
giant graviton spectrum computed by \cite{bhln,takayanagi}, of the
form $\frac{g_{YM}^2Nn^2}{J^2}$, are also of order $n^2/N$ in the
appropriate limit.

The AdS giants, for which our semiclassical considerations are most
reliable, are harder to analyze as operators on the field theory side.
Some of the difficulties are even apparent in the AdS dual.  For
example, it costs infinite energy to take these states to large
$\rho$, and they are not localized along the directions tangent to the
boundary, so it is not clear that there is a simple local operator
description in the dual field theory.  A proposal for the operators
corresponding to the AdS giants has been given in \cite{corley}; a
subsequent paper \cite{takayanagi} studied these operators in the
plane wave limit.

An alternative approach, in the spirit of the analysis of monopoles
and solitons \cite{dhn}, is to identify a classical field
configuration which solves the equations of motion and then to
consider quantum fluctuations about this solution.  The authors of
\cite{hashimoto} (whose argument we follow) found the relevant field
configuration for an $m$-dimensional field theory from the action for
the scalars
\beq
S=-\frac{1}{2g_{YM}^2} \int d^m x \left[(\partial \phi_1)^2+(\partial \phi_2)^2+\frac{(m-2)^2}{4L_{AdS}^2}(\phi_1^2 +\phi_2^2) \right].
\eeq
The fields $\phi_{1,2}$ are the scalars corresponding to the plane of
rotation of the expanded brane.  Because the geometry of the branes is
of the form $R\times S^{m-1}$, there is a mass term for the $\phi$
fields, whose form is fixed by conformal invariance.  Integrating over
the directions of the $S^{m-1}$, the action becomes
\beq
S=\frac{L^3 \Omega_{m-1}}{2 g_{YM}^2} \int dt \left[\dot{\phi_1}^2+\dot{\phi_2}^2-\frac{(m-2)^2}{4L_{AdS}^2}(\phi_1^2 +\phi_2^2) \right].
\label{ftscalars}
\eeq
The action is simply that of a two-dimensional harmonic oscillator, so the dynamics are exactly solvable.  We now set $L=1$ and make the change of variables
\beq
\phi_1 = \sqrt{\frac{g_{YM}^2N}{\Omega_{m-1}}}\eta \cos\theta, \qquad \phi_2 = \sqrt{\frac{g_{YM}^2N}{\Omega_{m-1}}}\eta \sin\theta
\label{etatheta}
\eeq
in terms of which the action is
\beq
S= \frac{N}{2} \int dt \left(\dot{\eta}^2+\eta^2 \dot{\theta}^2 -\frac{(m-2)^2}{4} \eta^2\right).
\eeq
The Hamiltonian and angular momentum $J=\frac{\delta S}{\delta \dot{\theta}}$ are conserved.  The energy is minimized at
\beq
\eta_0^2 = \frac{2J}{N(m-2)}
\eeq
and the spectrum is given by
\beq
E-\frac{m-2}{2} J = (m-2) n.
\eeq
This calculation reproduces the leading terms in
(\ref{specads5},\ref{specads4},\ref{specads7}).  

The effective action (\ref{ftscalars}) is accurate for perturbations
about a BPS configuration, but may receive large corrections for
non-BPS backgrounds, which are our main interest.  It is natural to
suspect that these corrections are responsible for the subleading
terms in the brane spectra computed earlier.  To see how these terms
arise in field theory, let us take $d=4$ for specificity, and note
that we can regard the brane configuration with a single AdS giant
graviton in the background of N static D3-branes as a theory with
$SU(N+1)$ gauge symmetry spontaneously broken to $SU(N)\times U(1)$.
The modes not in the light $SU(N)$ or $U(1)$ are Higgsed by the vacuum
expectation values of the $\phi$ fields.  When we integrate out these
massive modes there is a $\phi^4$ coupling induced at one-loop order,
which will scale as $1/N$.  We note that such a term, proportional to
$\frac{E-J}{E}$, was identified on the gravity side in \cite{d1d5},
and corresponded precisely to a non-cancellation of brane tension and
RR flux forces.

By considering a few simple one-loop Feynman diagrams, we can
schematically reproduce the $Nn^2/J^2$ dependence of the anomalous
dimension in (\ref{specads5}) in weakly coupled field theory.  A
direct comparison with the gravity result, which is valid at strong
coupling in the field theory, is of course impossible, so we will not
worry about the precise numerical factor or factors of $g_{YM}^2 N$.
Also, we will measure all dimensionful quantities in units of
$L_{AdS}$.  The relevant Feynman diagrams, shown in Figure 2, have
four external scalar legs; for definiteness we can consider the $U(1)$
component of the field $\eta$, which is described in terms of the $\phi$
fields in (\ref{etatheta}).
\begin{figure}
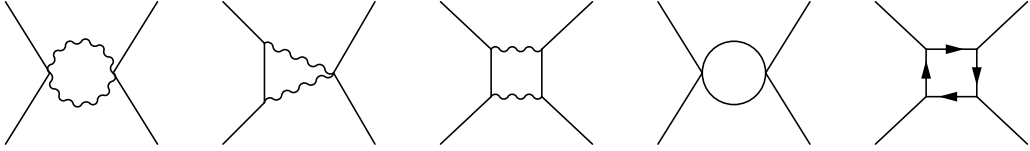

\begin{center}
$\begin{array}{ccccc}
\epsfxsize=1in
\epsffile{graph1.1} &
\epsfxsize=1in
\epsffile{graph2.1} &
\epsfxsize=1in
\epsffile{graph3.1} &
\epsfxsize=1in
\epsffile{graph4.1} &
\epsfxsize=1in
\epsffile{graph5.1} \\
\end{array}$
\end{center}
\caption{One loop diagrams which contribute to the four-scalar interaction.  Solid lines are scalars, directed lines are fermions, and wavy lines are vector mesons.}
\label{diagrams1}
\end{figure}
The internal lines are given by the Higgsed fields, which carry a
single $SU(N)$ index and have masses proportional to $\eta$.  In the
supersymmetric case, we have a constant $\eta=\eta_0=\sqrt{J/N}$, and
therefore constant masses.  In this case, the divergent part of the
one-loop amplitude cancels, leaving a finite piece of order $1/N$.  In
the semiclassical limit $n\gg 1$, this correction turns out to be
small compared to the non-supersymmetric contribution, and so we will
ignore it.  In the non-supersymmetric case, when the D3-branes are
oscillating, there is an additional finite piece which we cannot
ignore.  The most straightforward way to compute this correction is to
write $\eta = \eta_0 +\eta_1(t)$ and treat $\eta_1$ as a field with a
Yukawa-type coupling to the massive fields.  Thus, for example, the
Higgsed scalars have a coupling of the form
\beq
\half \eta^2 \phi^2 \sim \half \eta_0^2 \phi^2 + \eta_0\eta_1(t)\phi^2 
\eeq
and the associated vertex appears in Figure 3.
\begin{figure}
\begin{center}
\epsfxsize=.75in
\epsffile{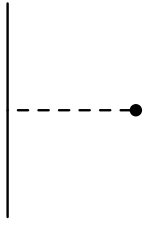}
\end{center}
\caption{Coupling of $\eta_1$ to the scalars.  There are analogous vertices for the vector particles and fermions.}
\end{figure}
Now we allow the $\eta$ field to fluctuate quantum-mechanically around
the classical solution and consider the four-scalar vertex.  Working
in momentum space, we hold the four fluctuating scalar legs at zero
momentum.  The loop diagrams with one insertion of $\eta_1$ do not
contribute because of momentum conservation, so the leading effect of
the time dependence arises from two $\eta_1$ insertions; a typical
Feynman diagram is shown in Figure 4.
\begin{figure}
\begin{center}
\epsfxsize=1in
\epsffile{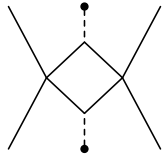}
\end{center}
\caption{A typical loop diagram with two $\eta_1$ insertions.} 
\end{figure}
Allowing these insertions to carry momentum $p$, a typical loop
integral will have the form
\beq
\hspace{-1.5cm}\frac{1}{N}\int d^4k \frac{\eta_0^2\eta_1(p)\eta_1(-p)}{(k^2+\eta_0^2)^2((k+p)^2+\eta_0^2)^2} &=& \frac{1}{N}\int d^4k \eta_0^2\eta_1(p)\eta_1(-p)\Bigg[\frac{1}{(k^2+\eta_0^2)^4} \nonumber \\ &&\hspace{2cm}-\frac{2p^2}{(k^2+\eta_0^2)^5} +\frac{12(p\cdot k)^2}{(k^2+\eta_0^2)^6} \Bigg] \\
&\sim& \frac{\eta_1(p)\eta_1(-p)p^2}{N\eta_0^4}.
\eeq
We have made the approximation that $p^2 \ll \eta_0^2$ in the top line
and retained the overall factor of $1/N$.  In the second line, we have
discarded the zero-order term in the $p^2$ expansion, which must
cancel amongst the one-loop diagrams.  Integrating over $p$ and
converting back to position space, we obtain the time-average
$\langle\dot{\eta}_1^2\rangle\sim n$ in the numerator, or
\beq
\frac{\langle \dot{\eta}_1^2\rangle}{N\eta_0^4} \sim \frac{Nn}{J^2}.
\eeq
Up to numerical and $g_{YM}^2 N$ factors, this is the one-loop
effective coupling for $\eta^4$, and including it in a semiclassical
computation gives a correction of order $Nn^2/J^2$ to the spectrum.

An important issue is that in our analysis we have ignored effects due
to backreaction of the probe D3-branes on the string background.
These effects are of order $1/N$, and may change our results.  One
effect of the backreaction, for example, is that the probe brane
should change the location of the center of mass of the system by an
amount of order $1/N$ -- essentially, there ought to be a reduced mass
shift of the spectrum.  Still, there should be a regime in which our
results are robust.  Backreaction effects like the reduced mass shift
should give a correction to the spectrum of order $n/N$.  On the field
theory side, for sphere giants, this might correspond to a $1/N$
correction to the bare masses of the $n$ impurities, or a combinatoric
correction ($\frac{n(n-1)}{2}$ rather than $n^2/2$ ways having
pairwise interactions, for example); for AdS giants there are
subleading finite contributions to the effective action after
integrating out massive fields.  Then, in the limit that $n \gg 1$ but
$n \ll N$, this backreaction effect is small compared to the $n^2/N$
contributions we computed above.
\footnote{We note, however, that in the case of M5-branes in $AdS_7$,
the semiclassical expansion proceeds in powers of $1/N^2$, so $1/N$
backreaction effects could actually be dominant.}  Clearly, a more
detailed computation of the backreaction in supergravity is necessary
to confirm or deny these handwaving arguments. Such a computation
might not be hopelessly ambitious, as it should eventually boil down
to a one-dimensional mechanics problem.

We have ignored various other effects in our calculation, in
particular the effects of massive string modes and of closed string
emission.  Notice that we have written the metric (\ref{metric}) in
such a way that all the factors of $g_s N$ factor out.  In the
subsequent analysis, no factors of $g_s N$ could appear.  However, we
considered only supergravity fields and not massive string modes,
which would have introduced $\alpha'/L^2 \sim 1/(g_s N)^{1/2}$
corrections.  Thus to ensure that we can ignore these effects, we must
also take $1/(g_sN)^{1/2} << n/N$.  As for closed string emission, at
leading order in $1/N$ these effects cannot change the spectrum
(though they do allow our non-BPS states to decay.)  Radiative
corrections to the spectrum due to closed strings (which roughly
correspond to non-planar diagrams in the field theory) only appear at
order $1/N^2$, and should not affect our results at order $1/N$.  To
summarize, we expect our semiclassical calculations for D3-branes in
$AdS_5 \times S^5$ to be sensible if
\beq
\frac{1}{N}, \frac{1}{(g_sN)^{1/2}} \ll \frac{n}{N} \ll 1.
\eeq

Though we have seen that semiclassical methods can give useful
information on both sides of the gauge/gravity duality, much remains
to be done.  On the gravity side, we need to understand backreaction
and stringy effects better, and on the field theory side, we need more
detailed computations of corrections to the effective action.  We also
hope that the semiclassical results for M-branes may give some hints
about the structure of the dual gauge theories on the boundaries of
$AdS_4$ and $AdS_7$.

\section*{Acknowledgments}

I am deeply grateful to Nissan Itzhaki and Igor Klebanov for
suggesting this problem, for reading the manuscript, and for many
helpful discussions.  I also thank Chris Beasley, John McGreevy, and
Sameer Murthy for useful conversations.  This material is based upon
work supported in part by the National Science Foundation under Grant
PHY-9802484.  Any opinions, findings, and conclusions or
recommendations expressed in this material are those of the authors
and do not necessarily reflect the views of the National Science
Foundation.


\begin{thebibliography}{99}
\bibitem{jthroat}
J.~Maldacena, ``The Large N limit of superconformal field theories and
supergravity,'' {\it Adv. Theor. Math. Phys.} {\bf 2} (1998) 231, 
{{\tt hep-th/9711200}}.

\bibitem{US}
S.S. Gubser, I.R. Klebanov, and A.M. Polyakov, ``Gauge theory correlators
from noncritical string theory,''
{\it Phys. Lett.} {\bf B428} (1998) 105,
{{\tt hep-th/9802109}}.

\bibitem{EW}
E.~Witten, ``Anti-de Sitter space and holography,''
{\it Adv. Theor. Math. Phys.} {\bf 2} (1998) 253,
{{\tt hep-th/9802150}}.

\bibitem{GKP}
S.~S.~Gubser, I.~R.~Klebanov and A.~M.~Polyakov,
``A semi-classical limit of the gauge/string correspondence,''
Nucl.\ Phys.\ B {\bf 636}, 99 (2002)
{\tt hep-th/0204051}.

\bibitem{Minahan}
J.~A.~Minahan,
``Circular semiclassical string solutions on AdS(5) x S(5),''
{\tt hep-th/0209047}.

\bi{mms}
J.M.Maldacena, J.~Michelson and A.~Strominger,
``Anti-de Sitter fragmentation,''
JHEP {\bf 9902}, 011 (1999)
{\tt hep-th/9812073}

\bi{d1d5}
N.~Seiberg and E.~Witten,
``The D1/D5 system and singular CFT,''
JHEP {\bf 9904}, 017 (1999)
{\tt hep-th/9903224}

\bibitem{giants}
J.~McGreevy, L.~Susskind and N.~Toumbas,
``Invasion of the giant gravitons from anti-de Sitter space,''
JHEP {\bf 0006}, 008 (2000)
{\tt hep-th/0003075}

\bi{grisaru}
M.~T.~Grisaru, R.~C.~Myers and O.~Tafjord,
``SUSY and Goliath,''
JHEP {\bf 0008}, 040 (2000)
`{\tt hep-th/0008015}

\bi{hashimoto}
A.~Hashimoto, S.~Hirano and N.~Itzhaki,
``Large branes in AdS and their field theory dual,''
JHEP {\bf 0008}, 051 (2000)
{\tt hep-th/0008016}

\bibitem{das}
S.~R.~Das, A.~Jevicki and S.~D.~Mathur,
``Vibration modes of giant gravitons,''
Phys. Rev. D {\bf 63}, 024013 (2001)
{\tt hep-th/0009019}

\bi{mikhailov}
A.~Mikhailov, 
``Giant gravitons from holomorphic surfaces,''
JHEP {\bf 0001}, 027 (2000)
{\tt hep-th/0010206}

\bi{kim}
J.~Y.~Kim and Y.~S.~Myung,
``Vibration modes of giant gravitons in the background of dilatonic D-branes,''
Phys. Lett. B {\bf 509}, 157 (2001)
{\tt hep-th/0103001}

\bi{takayanagi}
H.~Takayanagi and T.~Takayanagi,
``Notes on giant gravitons on PP-waves,''
{\tt hep-th/0209160}

\bibitem{mo}
J.~M.~Maldacena and H.~Ooguri, 
``Strings in AdS(3) and SL(2,R) WZW Model. I,''
J.\ Math.\ Phys.\ {\bf 42}, 2929 (2001)
{\tt hep-th/0001053};
J.~M.~Maldacena and H.~Ooguri, 
``Strings in AdS(3) and SL(2,R) WZW Model. III:Correlation Functions,''
Phys.\ Rev.\ D{\bf 65}, 106006 (2002)
{\tt hep-th/0111180}.

\bi{bbns}
V.~Balasubramanian, M.~Berkooz, A.~Naqvi and M.~J.~Strassler,
``Giant gravitons in conformal field theory,''
JHEP {\bf 0204}, 034 (2002)
{\tt hep-th/0107119}.

\bibitem{corley}
S.~Corley, A.~Jevicki and S.~Ramgoolam,
``Exact correlators of giant gravitons from dual N = 4 SYM theory,''
Adv.\ Theor.\ Math.\ Phys.\  {\bf 5}, 808 (2001)
{\tt hep-th/0111222}.

\bi{aharony}
O.~Aharony, Y.~E.~Antebi, M.~Berkooz and R.~Fishman,
``'Holey sheets': Pfaffians and subdeterminants as D-brane operators in  large N gauge theories,''
{\tt hep-th/0211152}.

\bibitem{bhk}
D.~Berenstein, C.~P.~Herzog, I.R.Klebanov,
``Baryon Spectra and AdS/CFT Correspondence,''
JHEP {\bf 0206},047 (2002)
{\tt hep-th/0202150}

\bibitem{wittenbar}
E.~Witten,
``Baryons in the 1/N Expansion,''
Nucl.\ Phys.\ {\bf B160}, 57, (1979).

\bibitem{bhln}
V.~Balasubramanian, M.~x.~Huang, T.~S.~Levi, A.~Naqvi.
``Open Strings from N=4 Super Yang-Mills,''
JHEP {\bf 0208},037 (2002)
{\tt hep-th/0204196}

\bi{BMN}
D.~Berenstein, J.~M.~Maldacena and H.~Nastase,
``Strings in flat space and pp waves from N = 4 super Yang Mills,''
JHEP {\bf 0204}, 013 (2002)
{\tt hep-th/0202021]}.
 
\bibitem{dhn}
R.~F.~Dashen, B.~Hasslacher, A.~Neveu,
``Nonperturbative Methods and Extended Hadron Models in Field Theory. 1. Semiclassical Functional Methods,''
Phys.\ Rev.\ {\bf D10}, 4114 (1974).

\end{thebibliography}
\end{document}